\documentclass[aps,twocolumn,showpacs,preprintnumbers,amsmath,amssymb]{revtex4}
\usepackage{graphicx}          % Include this line if your
\usepackage{graphicx}
\usepackage{graphicx}% Include figure files
\usepackage{dcolumn}% Align table columns on decimal point
\usepackage{bm}% bold math

\begin{document}

%\begin{frontmatter}

\title{Exact Relativistic Ideal Hydrodynamical Solutions\\
in (1+3)D with Longitudinal and Transverse Flows}

\author{Jinfeng Liao}\email{jliao@lbl.gov}

\author{Volker Koch}\email{vkoch@lbl.gov}

\address{Nuclear Science Division, Lawrence Berkeley National Laboratory, MS70R0319, 1 Cyclotron Road, Berkeley, CA 94720.}

%\begin{keyword}                        % Five to ten keywords,
%Relativistic Hydrodynamics, Flow, Relativistic Heavy Ion
%Collisions
%\end{keyword}                             % keyword list or with the
                                          % help of the Automatica
                                          % keyword wizard

% keywords here, in the form: keyword \sep keyword

% PACS codes here, in the form: \PACS code \sep code
%\begin{keyword}
%\end{keyword}

\begin{abstract}
A new method for solving relativistic ideal hydrodynamics in
(1+3)D is developed. Longitudinal and transverse radial flows are
explicitly embedded and the hydrodynamic equations are reduced to
a single equation for the transverse velocity field only, which is
much more tractable. As an application we use the method to find
analytically all possible solutions with power dependence on
proper time and transverse radius. Possible application to the
Relativistic Heavy Ion Collisions and possible generalizations of
the method are discussed.
\end{abstract}

%\end{frontmatter}

\maketitle

\section{Introduction}
\label{intro}

Relativistic hydrodynamics has wide applications in a variety of
physical phenomena, ranging from the largest scales such as in
cosmology and astrophysics\cite{cosmos_astro} to the smallest
scales such as in the relativistic nuclear
collisions\cite{Heinz:2009xj,Teaney:2009qa}. For an introduction
on the general formalism, see e.g.
\cite{Landau_book,Weinberg_book,Ollitrault:2007du,Gourgoulhon:2006bn}.

Recently there has been a remarkably successful application of
Relativistic Ideal Hydrodynamics (RIHD) to the description of the
space-time evolution of the hot dense QCD matter created in the
Relativistic Heavy Ion Collider (RHIC) experiments. In the
collisions of two relativistically moving heavy nuclei, a lot of
energy is deposited in a small volume which soon creates an
equilibrated system of high energy density with special initial
geometry: extremely thin in the beam direction $\hat z$ while in
the transverse plane $\hat x - \hat y$ it is of the size of the
nuclei. The space-time evolution at RHIC is characterized by fast
longitudinal expansion (longitudinal flow) and relatively slowly
developing transverse expansion (radial and elliptic flow). In
non-central collisions the created matter on the transverse plane
$\hat x - \hat y$ is initially anisotropic: such initial spatial
anisotropy leads to different pressure gradients and thus
different accelerations of the flow along different azimuthal
directions. The resulting anisotropic transverse flow velocity
eventually translates into the anisotropic azimuthal distribution
of the final particle yield which is represented by the
experimental observable called elliptic flow $v_2$ --- one of the
milestone measurements at RHIC \cite{Voloshin:2008dg}. The RIHD
model calculations\cite{Teaney:2000cw,Kolb_Heinz,Hirano} (and more
recently its extension to include viscous
corrections\cite{viscous_hydro}),
 performed with realistic
initial conditions and Equation of State (E.o.S) for RHIC, are
able to reproduce the elliptic flow data at
low-to-intermediate transverse momenta for almost all particle
species and for various centralities, beam energies and colliding
nuclei. These achievements of RIHD have been the basis for the
RHIC discovery that the matter being created is a strongly-coupled
nearly-perfect
fluid\cite{Shuryak_review,Gyulassy,Kharzeev:2009zb,Schaefer:2009dj}
with extremely short dissipative length. It has been suggested
\cite{Liao:2008pu,Liao:2006ry,Chernodub:2006gu,D'Alessandro:2007su,Ratti:2008jz}
that the microscopic origin could be due to the strong scattering
via Lorentz force between the electric and magnetic degrees of
freedom coexisting in the created matter, with the magnetic ones
ultimately connected with the mechanism of QCD confinement
transition.

The great success of RIHD at RHIC has also inspired considerable
interest in the formal aspects of relativistic hydrodynamics,
particularly in analytical solutions of the RIHD equations with an
emphasis on possible application to RHIC, see e.g.
\cite{Csorgo:2002bi,Nagy:2007xn,Bialas:2007iu,Pratt:2008jj,Wong:2008ex,Sinyukov:2004am,Borshch:2007uf,Koide:2008nw}.
The idea to use exact simple RIHD solutions to describe the
multi-particle production in high energy collisions dated back to
Landau and Khalatnikov \cite{Landau_Khalatnikov}. An important
solution came from Hwa and Bjorken's works\cite{Hwa,Bjorken}, i.e.
the rapidity boost invariant (1+1)D solution which is widely used
to describe the longitudinal expansion at RHIC. Many of the above
mentioned recent works
\cite{Nagy:2007xn,Bialas:2007iu,Pratt:2008jj,Wong:2008ex}
concentrate on finding (1+1)D solutions that give an alternative
description of the longitudinal expansion and a more realistic
(non-boost-invariant) multiplicity distribution over rapidity.

Despite the progress in solving RIHD in (1+1)D, it is quite
difficult to solve them in higher dimensions. To develop methods
and find solutions in a realistic (1+3)D setting with potential
application for RHIC remains an attracting but demanding task. In
this work, we will develop a new method to find solutions in
(1+3)D with both longitudinal and transverse flows. In Section
III, we will show how the method can reduce the hydrodynamics
equations to a single constraint equation for the transverse
velocity field only. Using the derived equation, we will find all
solutions with power-law dependence on proper time and transverse
radius in Section IV. The physical relevance of our results to
RHIC and possible generalizations will be discussed in Section V.
We will also include a brief introduction of RIHD in Section II
and an illustration of the method in (1+1)D in the Appendices B
and C.

\section{Review of Relativistic Ideal Hydrodynamics in (1+3)D }
\label{sec:1}

The hydrodynamics equations in general are simply the conservation
laws for energy and momenta, i.e.
\begin{equation}\label{eqn_hydro}
T^{m n}\,  _{; \, n}=0
\end{equation}
with $m,n$ running over $3+1$ space-time indices. Following usual
convention (in e.g. \cite{Landau_book,Weinberg_book}), the
subscript ``$;\, n$'' denotes the covariant derivative $D_n$ while a
subscript ``$,\, n$'' is for ordinary derivative $\partial_n$.
Below we will introduce curved coordinates in order to simplify
the hydrodynamics equations. Therefore, we use the general form
for the hydrodynamics equations which involves covariant
derivatives, see e.g. \cite{Landau_book,Weinberg_book}. Throughout
this paper we discuss only hydrodynamics without any  conserved
charge, leaving the situation with conserved currents for further
investigation.

For relativistic ideal hydrodynamics, the stress tensor is given
by
\begin{equation} \label{eqn_tmunu}
T^{m n} = (\epsilon + p)\, u^m u^n - p \, g^{m n}
\end{equation}
with $\epsilon,p$ the energy density and pressure defined in the
flowing matter's local rest frame (L.R.F) which by definition are
Lorentz scalars. The flow field $u^{m}(x)$ is constrained by
$u^m\cdot u_m =1$. In the usual $(t,\vec x)$ coordinates one can
express $u^{m}(x)$ as $\gamma (1,\vec v)$ with
$\gamma=1/\sqrt{1-\vec v^2}$ and $\vec v=d\vec x/dt$.

We further need to specify an equation of state (E.o.S) relating
the energy density $\epsilon$ and the pressure $p$ of the
underlying fluid. Here we employ a simple, linear E.o.S, which is
typically used in analytic studies of RIHD
\cite{Nagy:2007xn,Bialas:2007iu,Pratt:2008jj,Wong:2008ex,Hwa,Bjorken}:
\begin{equation} \label{eqn_eos}
 p=\nu (\epsilon+p)
\end{equation}
The above means $\epsilon=\frac{1-\nu}{\nu} p$ implying a speed of
sound $c_s=\sqrt{\partial p \over
\partial \epsilon}=\sqrt{\frac{\nu}{1-\nu}}$, and, in order to assure $c_s\le 1$,
we require $0<\nu\le 1/2$. We note, that by adding a constant to
the above E.o.S, the resulting velocity field remains unchanged.

The hydrodynamics equations together with the E.o.S thus form a
complete set of 5 equations for the 5 field variables:
$\epsilon(x),p(x)$ and the three independent components of
$u^m(x)$.

\subsection{Hydro Equations in Curved Coordinates}

When formulating hydrodynamics for application to e.g. the
relativistic heavy ion collisions, it is often useful to use
alternative coordinates systems which are curved. For our purpose
of studying the (1+3)D solutions with longitudinal and transverse
flow, we will use a coordinate system of $(\tau,\eta,\rho,\phi)$:
i.e. the proper time, the spatial (longitudinal) rapidity, the
transverse radius and the azimuthal angle. They are related to the
usual $(t,x,y,z)$ in the following way:
\begin{eqnarray}
&& \tau=\sqrt{t^2-z^2} \, , \,
\eta=\frac{1}{2}\mathbf{ln}\frac{t+z}{t-z}\, ,
\nonumber \\
&& \rho=\sqrt{x^2+y^2} \, , \,
\phi=\frac{1}{2i}\mathbf{ln}\frac{x+y\cdot i}{x-y\cdot i}
\end{eqnarray}
and inversely
\begin{eqnarray}
&& t=\tau \cosh \eta \, , \,  z=\tau \sinh \eta \, , \,
\nonumber \\
&& x=\rho \cos\phi \, , \, y=\rho \sin\phi
\end{eqnarray}
The velocity field $u^{m}$ in these coordinates is related to
$u^{\mu}=\gamma(1,\vec v)$ in flat coordinates $(t,\vec x)$ via
\begin{eqnarray} \label{eqn_curved_v}
&& u^\tau = \gamma ( \cosh \eta -v_z\sinh \eta) \, , \,  u^\eta =
\frac{\gamma}{\tau} ( v_z \cosh \eta -\sinh \eta)  \, , \,
\nonumber \\
&& u^\rho = \gamma ( v_x \cos \phi + v_y \sin \phi) \, , \, u^\phi
= \frac{\gamma}{\rho} (v_y \cos \phi - v_x \sin \phi) . \nonumber \\
\end{eqnarray}

The metric tensor associated with the $(\tau,\eta,\rho,\phi)$
coordinates is
\begin{eqnarray} \label{eqn_metric}
&& g_{mn}=Diag(1,-\tau^2,-1,-\rho^2) \nonumber \\
&& g^{mn}=Diag(1,-\frac{1}{\tau^2},-1,-\frac{1}{\rho^2})
\end{eqnarray}
For the covariant derivatives we will need the Affine connections
$\Gamma^{j}_{mn}= g^{jk}\Gamma_{kmn}=g^{jk} \frac{1}{2}(g_{km}\,
_{,\, n}+g_{kn}\, _{,\, m}-g_{mn}\, _{,\, k})$. In our case, the
non-vanishing connections are:
\begin{eqnarray} \label{eqn_connection}
\Gamma^\tau_{\eta \eta}=\tau \, , \, \Gamma^{\eta}_{\eta
\tau}=\Gamma^{\eta}_{\tau \eta}=\frac{1}{\tau}
\nonumber \\
\Gamma^{\rho}_{\phi \phi}=-\rho \, , \, \Gamma^{\phi}_{\rho
\phi}=\Gamma^{\phi}_{\phi \rho}=\frac{1}{\rho}
\end{eqnarray}
We also give the explicit form of covariant derivatives in the
present coordinates for an arbitrary contra-variant-vector $A^k$
(i.e. with upper indices $k=\tau,\eta,\rho,\phi$) :
\begin{eqnarray} \label{eqn_covariant_d}
A^k\, _{;\, \tau} &=& A^k\, _{,\, \tau} + \Gamma^k_{\tau\, i}A^i =
A^k\, _{,\, \tau} + \frac{1}{\tau}\, \delta^k_\eta \, A^\eta
\nonumber \\
A^k\, _{;\, \eta} &=& A^k\, _{,\, \eta} + \Gamma^k_{\eta\, i}A^i =
A^k\, _{,\, \eta} + \tau\, \delta^k_\tau \, A^\eta +
\frac{1}{\tau}\, \delta^k_\eta \, A^\tau
\nonumber \\
A^k\, _{;\, \rho} &=& A^k\, _{,\, \rho} + \Gamma^k_{\rho\, i}A^i =
A^k\, _{,\, \rho} + \frac{1}{\rho}\, \delta^k_\phi \, A^\phi
\nonumber \\
A^k\, _{;\, \phi} &=& A^k\, _{,\, \phi} + \Gamma^k_{\phi\, i}A^i =
A^k\, _{,\, \phi} - \rho\, \delta^k_\rho\, A^\phi +
\frac{1}{\rho}\, \delta^k_\phi \, A^\rho
\nonumber \\
\end{eqnarray}
Inserting the above
Eqs.(\ref{eqn_tmunu},\ref{eqn_curved_v},\ref{eqn_metric}) into the
general hydrodynamics equations (\ref{eqn_hydro}) and making use
of Eq.(\ref{eqn_connection},\ref{eqn_covariant_d}), one obtains
the hydrodynamics equations explicitly in the curved coordinates
$(\tau,\eta,\rho,\phi)$.

\subsection{Some Known Simple Exact Solutions}
\label{sec:2.2}

We now recall some known simple exact solutions that are pertinent
for our approach.

One of the most famous examples is the so-called Hwa-Bjorken
solution \cite{Hwa,Bjorken} which is essentially the Hubble expansion in (1+1)D. The
pressure and velocity fields of this solution are given by
\begin{eqnarray} \label{eqn_bj}
&& p_{Bj.} =  \frac{constant}{\tau^{1/(1-\nu)}}  \nonumber \\
&& u_{Bj.}=(1 , 0 , 0 , 0)
\end{eqnarray}
It is more transparent to look at the components of $\vec v$ in
flat coordinates, which are simply $v_z=\tanh \eta=\frac{z}{t} \,
, \, v_x=v_y=0$.

A generalization of the Hwa-Bjorken solution to radial Hubble flow
in (1+3)D (with further generalization to (1+d)D ) is
straightforward. The pressure and velocity fields are
\begin{eqnarray} \label{eqn_hu}
&& p_{Hu.} = \frac{constant}{(\tau^2-\rho^2)^{\frac{3}{2(1-\nu)}}}
\nonumber \\
&& u_{Hu.}=\gamma (\frac{1}{\cosh\eta} , 0 , \frac{\rho}{\tau} ,
0) \, , \,
\gamma=\frac{\cosh\eta}{\sqrt{1-(\rho/\tau)^2\cosh^2\eta}}
\nonumber \\
\end{eqnarray}
In flat coordinates the velocity fields are given in simple form:
$\vec v =\vec x/t = (x/t,y/t,z/t)$.

\section{The New Reduction Method}

In this section, we use a new reduction method to find solutions
for (1+3)D RIHD equations. The general idea is to first embed
known solutions in lower dimensions which automatically solve 2
out of the total of 4-component hydrodynamics equations, and then
reduce the remaining 2 equations into a single equation for the
velocity field only. As usual, one starts with a certain ansatz
for the flow velocity field: in our case we will use an ansatz
with built-in longitudinal and transverse radial flow, aiming at
possible application for RHIC. It would be even more interesting
to include {\em transverse elliptic flow} which requires a
suitable curved coordinates (like certain hyperbolic coordinates)
other than the one used here. However generally in those cases,
more Affine connections are non-vanishing, which makes the
reduction method discussed below much more involved: we will leave
this for future investigation.

\subsection{Including Longitudinal and Transverse Flow}

We first embed the boost-invariant longitudinal flow as many
numerical hydrodynamics calculations do, which is a suitable
approach for RHIC related phenomenology. To do that, we simply set
$v_z=z/t=\tanh \eta$, i.e. $u^{\eta}=0$.

Next we  include the transverse radial flow which is isotropic in
the transverse plane. Radial flow is substantial and important at
RHIC.  To do so, we introduce the radial flow field $v_\rho$ and
set the flat-coordinate transverse flow fields to be $v_x=v_\rho
\cos\phi$ and $v_y=v_\rho \sin\phi$, which implies for the curved
coordinates $u^\rho=\gamma v_\rho$ and $u^\phi=0$. We note that
this ansatz goes beyond a simple change to cylindrical
coordinates, since we require that $u^\phi=0$ which considerably
simplifies the hydrodynamics equations.

To summarize, in order to describe a situation with both
longitudinal flow and transverse radial flow we have made the
following ansatz for the flow fields $u^m$ in the coordinates
$(\tau,\eta,\rho,\phi)$:
\begin{eqnarray} \label{eqn_flow_embed}
&& u^m=\bar\gamma \, (\, 1, \, 0, \, \bar v_\rho, \, 0)  \\
&& \bar v_\rho \equiv v_\rho \cosh\eta  \quad , \quad \bar\gamma
\equiv 1/\sqrt{1-\bar v_\rho^2} \nonumber
\end{eqnarray}
Note that we need to require $\bar v_\rho \le 1$.

\subsection{The Equation for Transverse Velocity}

With the flow fields given in (\ref{eqn_flow_embed}), we can now
explicitly express the stress tensor components. The non-vanishing
ones are given below:
\begin{eqnarray}
&&
T^{\tau\tau}=\bar\gamma^2(\epsilon+p)-p=(\frac{\bar\gamma^2}{\nu}-1)
p \\
&& T^{\rho\rho}=\bar\gamma^2\bar
v_{\rho}^2(\epsilon+p)+p=(\frac{\bar\gamma^2\bar v_\rho^2}{\nu}+1) p\\
&& T^{\tau\rho}=\bar\gamma^2\bar v_\rho (\epsilon+p)=\frac{\bar
\gamma^2\bar v_\rho}{\nu} p \\
&& T^{\eta\eta}=\frac{p}{\tau^2} \quad , \quad
T^{\phi\phi}=\frac{p}{\rho^2}
\end{eqnarray}
For the second equalities in each of the first three lines we have
used the E.o.S (\ref{eqn_eos}) to substitute $\epsilon+p$ by
$p/\nu$.

With the above expressions and using
(\ref{eqn_connection})(\ref{eqn_covariant_d}), the hydrodynamics
equations (\ref{eqn_hydro}) then become
\begin{eqnarray}
\label{eqn_reduced_a} && T^{\tau\lambda} \,
_{;\,\lambda}=T^{\tau\tau} \, _{,\,\tau}+\frac{T^{\tau\tau}}{\tau}
+\frac{p}{\tau}+T^{\tau\rho} \,
_{,\,\rho}+\frac{T^{\tau\rho}}{\rho}=0 \quad \\
&& T^{\eta\lambda} \, _{;\,\lambda}=\frac{1}{\tau^2} p \, _{,
\,\eta}=0 \label{eqn_eta} \\
\label{eqn_reduced_b} && T^{\rho\lambda} \,
_{;\,\lambda}=T^{\rho\rho} \, _{,\,\rho}+\frac{T^{\rho\rho}}{\rho}
-\frac{p}{\rho}+T^{\tau\rho} \,
_{,\,\tau}+\frac{T^{\tau\rho}}{\tau}=0 \quad \\
&& T^{\phi\lambda} \, _{;\,\lambda}=\frac{1}{\rho^2} p \, _{,
\,\phi}=0 \label{eqn_phi}
\end{eqnarray}
The two equations involving derivatives over $\eta$ and $\phi$ are
trivially solved by setting $p(x)=p(\tau,\rho)$ (and the same for
energy density $\epsilon(\tau,\rho)$ due to the E.o.S) and
accordingly $\bar v_\rho (x)= \bar v_\rho (\tau,\rho)$. We note,
that the simple form of Eqs.(\ref{eqn_eta},\ref{eqn_phi}), are a direct
consequence of the vanishing components $u^\eta=u^\phi=0$ in the
flow field ansatz, Eq.(\ref{eqn_flow_embed}).

Finally we introduce a combined field variable ${\mathcal K}$
defined as
\begin{equation}
{\mathcal K}\equiv \frac{T^{\tau\tau}+p}{(\rho \tau)} =\frac{
 \bar\gamma^2 p}{\nu \rho \tau} \,\, \to \,\,  p=
\frac{\nu \rho \tau }{\bar\gamma^2} {\mathcal K}
\end{equation}
We then substitute the pressure $p$ in the equations
(\ref{eqn_reduced_a}) and (\ref{eqn_reduced_b}) and
obtain two equations for the fields ${\mathcal K}$ and $\bar
v_\rho$, which can be  expressed as
\begin{eqnarray} \label{eqn_S}
&& \label{eqn_S_1} D_a \cdot {\mathcal K} _{, \, \tau} + D_b \cdot {\mathcal K} _{, \, \rho}= D_1 \cdot {\mathcal K} \\
&& \label{eqn_S_2} D_b \cdot {\mathcal K} _{, \, \tau} + D_c \cdot
{\mathcal K} _{, \, \rho}= D_2 \cdot {\mathcal K}
\end{eqnarray}
The coefficients $D_a,D_b,D_c,D_1,D_2$ are given by:
\begin{eqnarray}
&& D_a=(1-\nu) + \nu \bar v_\rho^2 \nonumber \\
&& D_b= \bar v_\rho \nonumber \\
&& D_c= \nu + (1-\nu) \bar v_\rho^2 \nonumber \\
&& D_1= -2\nu {\bar v_\rho} {\bar v_\rho}\, _{, \, \tau} - \nu
(1-{\bar v_\rho}^2)/\tau
- {\bar v_\rho}\, _{, \, \rho} \nonumber \\
&& D_2= -2(1-\nu)\bar v_\rho {\bar v_\rho}\, _{, \, \rho}
+ \nu (1-{\bar v_\rho}^2)/\rho - \bar v_\rho \, _{, \, \tau}  \nonumber \\
\end{eqnarray}

From (\ref{eqn_S_1})(\ref{eqn_S_2}) we  obtain
\begin{eqnarray} \label{eqn_lnS}
&& \frac{{\mathcal K}\, _{, \, \tau}}{\mathcal K} =
(\mathbf{ln}\,{\mathcal K})\, _{, \, \tau} =
\frac{D_c D_1-D_b D_2}{D_a D_c- D_b^2} \equiv {\mathcal F}[\tau,\rho] \nonumber \\
&& \frac{{\mathcal K}\, _{, \, \rho}}{\mathcal K} =
(\mathbf{ln}\,{\mathcal K})\, _{, \, \rho} = \frac{D_a D_2-D_b
D_1}{D_a D_c -D_b^2} \equiv {\mathcal G}[\tau,\rho] \nonumber
\\
\end{eqnarray}
with the functions $\mathcal F$,$\mathcal G$ given by
\begin{eqnarray} \label{eqn_f}
&& {\mathcal F}[\bar v_\rho(\tau,\rho)]= \frac{1}{\nu (1-\nu)(1-\bar v_\rho^2)^2} \times \nonumber \\
&& {\bigg \{ }  [(1-\nu)\bar v_\rho^2-\nu]\bar v_\rho \, _{,\,
\rho} + \nonumber \\
&& \quad [(1-2\nu^2)+2\nu(\nu-1)\bar v_\rho^2] \bar v_\rho \bar v_\rho \, _{,\, \tau} \nonumber \\
&& \, -\nu [1-\bar v_\rho^2][\nu+(1-\nu)\bar v_\rho^2]/\tau -
\nu \bar v_\rho [1-\bar v_\rho^2]/\rho {\bigg \} }  \\
&&\label{eqn_g} {\mathcal G}[\bar v_\rho(\tau,\rho)]= \frac{1}{\nu (1-\nu)(1-\bar v_\rho^2)^2} \times \nonumber \\
&& {\bigg \{ }  [(\nu\bar v_\rho^2+(\nu-1)]\bar v_\rho \, _{,\,
\tau} + \nonumber \\
&& \quad
[(-1+4\nu-2\nu^2)+2\nu(\nu-1)\bar v_\rho^2] \bar v_\rho \bar v_\rho \, _{,\, \rho} \nonumber \\
&&  +\nu \bar v_\rho [1-\bar v_\rho^2]/\tau + \nu [1-\bar
v_\rho^2] [(1-\nu)+\nu \bar v_\rho^2]/\rho {\bigg \} }
\end{eqnarray}

In (\ref{eqn_lnS}), the function $\mathbf{ln}\,{\mathcal K}$
depends (via $\bar v_\rho$) on two variables $\tau$ and $\rho$,
and we have two equations for the two first order derivatives
$\frac{\partial \mathbf{ln}\,{\mathcal K}}{\partial \tau}$ and
$\frac{\partial \mathbf{ln}\,{\mathcal K}}{\partial \tau}$. For
$\mathbf{ln}\,\mathcal K$ as a single function of two variables
$\tau ,\rho$, the two equations can be consistent only if the
following constraint on second order derivatives are satisfied
$\frac{\partial^2 Ln\,{\mathcal K}}{\partial \tau
\partial \rho}=\frac{\partial^2 Ln\,{\mathcal K}}{\partial \rho \partial
\tau}$, i.e.
\begin{equation} \label{eqn_v_hydro}
\frac{\partial}{\partial \rho}{\mathcal F}[\bar v_\rho (\tau,\rho)] -
\frac{\partial}{\partial \tau}{\mathcal G}[\bar v_\rho (\tau,\rho)]=0
\end{equation}
Thus we only need to solve the above single equation for the
velocity field $\bar v_\rho(\tau,\rho)$. Since ${\mathcal
F},{\mathcal G}$ already involve the first derivatives of $\bar
v_\rho\, _{,\, \tau}$ and $\bar v_\rho\, _{,\, \rho}$, the
reduced velocity equation  \ref{eqn_v_hydro} is a second-order partial
differential equation for the velocity field. As a minor caveat,
the method applies to the case/region in which
$\mathbf{ln}\,\mathcal K$ is at least second-order differentiable.
This reduction method can be demonstrated in the more explicit
case of (1+1)D hydrodynamics, see Appendices B and C.

Given the above constraints, we can then solve from
(\ref{eqn_lnS}) the matter field $S$ directly
\begin{eqnarray}
{\mathcal K} ={\mathcal K}_0 \cdot e^{ {\big [} \int_{\tau_0}^\tau
d\tau' {\mathcal F[\tau',\rho]} + \int_{\rho_0}^\rho d\rho'
{\mathcal G[\tau,\rho']} {\big ]} }
\end{eqnarray}
with ${\mathcal K}_0$ being the value at arbitrary reference point
$\tau_0 \, , \,  \rho_0$.

Finally let us summarize our approach: after including into the
flow field ansatz the physically desired longitudinal and
transverse flows, we have reduced the hydrodynamic equations into
a single equation (\ref{eqn_v_hydro}) involving {\em ONLY} the
transverse velocity field $\bar v_\rho$, and any solution to this
equation automatically leads to the pressure field which together
with the velocity field forms a solution to the original
hydrodynamics equations:
\begin{eqnarray} \label{eqn_reduced_p}
p =constant  \times \frac{\rho\tau}{\kappa \bar \gamma^2} \times
e^{ {\big [} \int^\tau d\tau' {\mathcal F[\tau',\rho]} + \int^\rho
d\rho' {\mathcal G[\tau,\rho']} {\big ]} }
\end{eqnarray}

\subsection{Examination of the Method}

We now examine the correctness of the reduced equation
(\ref{eqn_v_hydro}) and the solution (\ref{eqn_reduced_p}), using
the two known simple analytic solutions (\ref{eqn_bj}) and
(\ref{eqn_hu}) as both of them are certain special cases of our
embedding with longitudinal and transverse radial flows.

For the 1-D Bjorken expansion, we have $\bar v_\rho \, _{Bj.}=0$
which leads to
\begin{eqnarray} \label{eqn_check_bj}
{\mathcal F}_{Bj.}=\frac{\nu}{\nu-1}\frac{1}{\tau} \quad , \quad
{\mathcal G}_{Bj.}=\frac{1}{\rho}
\end{eqnarray}
One can easily verify that the above ${\mathcal F}_{Bj.},{\mathcal
G}_{Bj.}$ satisfy the reduced equation (\ref{eqn_v_hydro}).
Furthermore by inserting ${\mathcal F}_{Bj.},{\mathcal G}_{Bj.}$
into the solution (\ref{eqn_reduced_p}) one finds exactly the
pressure in (\ref{eqn_bj}).

For the 3-D Hubble expansion, we have $\bar v_\rho \,
_{Hu.}=\rho/\tau$ which leads to
\begin{eqnarray} \label{eqn_check_hu}
&& {\mathcal
F}_{Hu.}=\frac{3}{\tau}+\frac{\nu-5/2}{1-\nu}\frac{2\tau}{\tau^2-\rho^2}
\nonumber \\
&& {\mathcal
G}_{Hu.}=\frac{1}{\rho}+\frac{\nu-5/2}{1-\nu}\frac{-2\rho}{\tau^2-\rho^2}
\end{eqnarray}
Again it can easily shown that the above ${\mathcal
F}_{Hu.},{\mathcal G}_{Hu.}$ satisfy the reduced equation
(\ref{eqn_v_hydro}). Furthermore by inserting ${\mathcal
F}_{Hu.},{\mathcal G}_{Hu.}$ into the solution
(\ref{eqn_reduced_p}) one finds exactly the pressure in
(\ref{eqn_hu}).

\section{Application of the Method}

As an example for an application of the embedding-reduction method in
the previous section, we show how to find all possible solutions
with the following ansatz for the radial velocity field:
\begin{equation} \label{eqn_app_v}
\bar v_\rho= A\cdot \tau^B \cdot \rho^C
\end{equation}
with $A,B,C$ arbitrary real numbers. We note that the two known
exact solutions we mentioned are special cases of the above form:
the 1-D Bjorken expansion corresponds to $A=0$ while the 3-D
Hubble expansion corresponds to $A=1,B=-1,C=1$. The velocity field
(\ref{eqn_app_v}), when put into (\ref{eqn_f})(\ref{eqn_g}), gives
the following
\begin{eqnarray}
&& \label{eqn_3d_f} {\mathcal F}[\tau,\rho]= \frac{1-2B}{\tau}-\frac{B+(1-4B)\nu+2B\nu^2}{\nu(1-\nu)}
\cdot \frac{1}{\tau[1-\bar v_\rho^2]} \nonumber \\
&& -\frac{C+(1-C)\nu}{\nu(1-\nu)}\cdot \frac{\bar v_\rho}{\rho [1-\bar v_\rho^2]}
+ \frac{B(1-2\nu)}{\nu(1-\nu)}\cdot \frac{1}{\tau [1-\bar v_\rho^2]^2}   \nonumber \\
&& +\frac{C(1-2\nu)}{\nu(1-\nu)}\cdot \frac{\bar v_\rho}{\rho [1-\bar v_\rho^2]^2} \\
&& \label{eqn_3d_g} {\mathcal G}[\tau,\rho]=
\frac{-2C-\nu/(1-\nu)}{\rho} - \frac{2C\nu^2-\nu-C}{\nu(1-\nu)}\cdot \frac{1}{\rho [1-\bar v_\rho^2]} \nonumber  \\
&& -\frac{B-1}{1-\nu}\cdot \frac{\bar v_\rho}{\tau [1-\bar v_\rho^2]}
+\frac{C(2\nu-1)}{\nu (1-\nu)}\cdot \frac{1}{\rho [1-\bar v_\rho^2]^2}  \nonumber \\
&& + \frac{B(2\nu-1)}{\nu (1-\nu)}\cdot \frac{\bar v_\rho}{\tau
[1-\bar v_\rho^2]^2}
\end{eqnarray}
We have used $\bar v_\rho\, _{,\, \rho}=\bar v_\rho \cdot C /
\rho$ and $\bar v_\rho\,  _{,\, \tau}= \bar v_\rho \cdot B /
\tau$. As a check of the above result, one can verify that by
setting $A=0$ they reduce to (\ref{eqn_check_bj}) while by setting
$A=1,B=-1,C=1$ they reduce to (\ref{eqn_check_hu}), as they
should.

By inserting (\ref{eqn_3d_f})(\ref{eqn_3d_g}) into
equation (\ref{eqn_v_hydro}), one obtains a rather complicated
constraint equation for the constants $A,B,C$. However after
a lengthy calculations, all possible combinations of
$A,B,C$ solving the equation can actually be exhausted. Leaving
the detailed (and technical) derivations to the Appendix A, we
only list the final results here:
\begin{itemize}
\item \textbf{Solution-I}: $A=0$ with $0<\nu\le \frac{1}{2}$ (1-D
Bjorken)
--- see (\ref{eqn_bj});\newline

\item \textbf{Solution-II}: $A=1,B=-1,C=1$ with $0<\nu\le
\frac{1}{2}$ (3-D Hubble) --- see (\ref{eqn_hu});\newline

\item \textbf{Solution-III}: $A=1,B=1,C=-1$ with $0<\nu\le
\frac{1}{2}$ --- the solutions are
\begin{eqnarray}
&& v_x=\frac{x}{t}\cdot \frac{t^2-z^2}{x^2+y^2} \, , \,
v_y=\frac{y}{t}\cdot \frac{t^2-z^2}{x^2+y^2} \, , \,
 v_z=\frac{z}{t} \nonumber \\
&&  p = \frac{constant}{(\tau\rho)^{1/(1-\nu)}
  (\rho^2-\tau^2)^{(1-3\nu)/(2\nu-2\nu^2)}}
\end{eqnarray}

\item \textbf{Solution-IV}: $A=1,B=1/3,C=-1/3$ with $\nu=1/4$ ---
the solutions are
\begin{eqnarray}
&& v_x=\frac{x}{t}\cdot {\big (}\frac{t^2-z^2}{x^2+y^2} {\big
)}^{2/3} \, , \, v_y=\frac{y}{t}\cdot {\big
(}\frac{t^2-z^2}{x^2+y^2} {\big )}^{2/3} \, , \,
 v_z=\frac{z}{t} \nonumber \\
&& p = constant\times
\frac{(\rho^{2/3}-\tau^{2/3})^{2/3}}{(\rho\tau)^{4/3}}
\end{eqnarray}

\item \textbf{Solution-V}: $A=-1,B=-1,C=1$ with $\nu=1/2$ --- the
solutions are
\begin{eqnarray}
&& v_x=\frac{-x}{t} \, , \, v_y=\frac{-y}{t} \, , \,
 v_z=\frac{z}{t} \nonumber \\
&& p = constant\times (\tau^2-\rho^2)
\end{eqnarray}

\end{itemize}
It can be verified that these solutions obtained by the method
introduced here are indeed solutions of the original hydrodynamics
equations (\ref{eqn_hydro}). One should notice the different
applicable kinematic regions in each of the above solutions which
comes from the constraint that the flow velocity shall be less
than the speed of light. For our solutions, the constraint is
$z\le t\,\, \& \,\, \sqrt{x^2+y^2+z^2}\le t $ for Solution-I,II,V
, and for Solution-III,IV the constraint is $z\le t\,\, \& \,\,
\sqrt{x^2+y^2+z^2}\ge t $.  For a detailed discussion about
solutions in different regions with respect to kinematic light
cone, see e.g. appendices of \cite{Nagy:2007xn}.

We notice that all the solutions (except the trivial Solution-I
with $A=0$) satisfy two features (1) $B=-C$ and (2) $|A|= 1$. The
first feature may be due to dimensional reasons. The second
feature, $|A|= 1$, may be heuristically understood in the
following way. We first consider the case $B=-C<0$, i.e. $\bar
v_\rho = A (\rho/\tau)^C$ with $C>0$: in this case the solution
exists in the region $\rho\le \tau\cdot |A|^{-1/C}$, and in
particular $\rho=0$ for $\tau=0$. Thus for any $\tau > 0$, the
flow front which travels with the speed of light, $|\bar
v_\rho|=1$, is located at $\rho=\tau$ and hence, $|A|=1$. Next we
consider the case $B=-C>0$, i.e. $\bar v_\rho = A (\tau/\rho)^B$
with $B>0$: in this case the solution exists in the region $\rho >
\tau \cdot A^{1/B}$ with $A>0$, separated from an empty region by
the boundary at $\rho = \tau \cdot A^{1/B}$. At this boundary, the
flow velocity approaches the speed of light $\bar v_\rho \to 1$
which enforces the matter density to drop to zero in order to
avoid an infinite $T^{mn}$ (due to the $\gamma$-factor in
Eq.(\ref{eqn_tmunu})). We imagine that at time $\tau=0$ the matter
fills the whole space and then starts to flow outward, thus the
boundary also moves outward from the origin with the speed of
light $\bar v_\rho=1$. This again implies the boundary should lie
at $\rho=\tau$ requiring  $A=1$.

The above example of the proposed embedding-reduction method
demonstrates the advantage of analytical solutions. Not only could
we find some solutions of the specific type (\ref{eqn_app_v}) but
we actually were able to {\em exhaust} all solutions of this type.
This also implies that for parametrization of flow velocity field,
like e.g. in the blast wave model for RHIC fireball, there are
only very limited choices for the flow profile ansatz.

\section{Summary and Discussion}

In summary, a general framework for the analytical treatment of
RIHD equations has been developed. The method features a
separation of longitudinal and transverse expansions, as inspired
by RHIC phenomenology. After the separation, the longitudinal and
transverse radial flows are embedded utilizing lower-dimensional
solutions. The remaining equations are found to be reducible to a
single constraint equation for transverse radial flow velocity
field only, which can be solved completely for a certain ansatz for the velocity field.
All solutions with power-law dependence on proper time and
transverse radius have been found.

We now discuss various possible extensions of the present
approach.

{\em Nontrivial longitudinal embedding:} In the current work the
longitudinal flow is embedded with the Hwa-Bjorken solution. It
would be very interesting to try embedding the newly found (1+1)D
solutions in
e.g.\cite{Nagy:2007xn,Bialas:2007iu,Pratt:2008jj,Wong:2008ex} with
more realistic longitudinal expansion for RHIC which would be
useful for studying elliptic flow in the forward/backward
rapidity and their correlation\cite{Liao:2009ni}.

{\em Solutions with non-power-law transverse expansion:} It would
also be interesting to a test more nontrivial ansatz for the
embedded transverse flow. For example we know from numerical
calculations of radial flow \cite{Kolb:2003gq} in central
collisions at RHIC that the radial velocity field may be
parameterized as $v_\rho\approx f(\tau) r/\tau$ with
$f(\tau\to0)\to 0$ and $f(\tau>>1 )\to 1$. Such parametrization
can be cast into the derived velocity equation (\ref{eqn_v_hydro})
to find possible solutions.

{\em Small deformation and elliptic flow:} The analytic treatment
of transverse elliptic flow is difficult. One approximate method
may be to introduce a parametrically small deformation of the
matter field (with a certain eccentricity parameter $\epsilon_2$)
on top of an exact solution with transverse radial flow and using
linearized hydrodynamics equations to investigate possible
universal relations between the finally developed velocity field
anisotropy $v_2$ and the initial $\epsilon_2$
\cite{Ollitrault:1992bk}.

{\em Transverse elliptic flow embedding:} Another possibility to
seek exact solutions with transverse elliptic flow is to use
instead of $(\rho,\phi)$  certain hyperbolic coordinates which
by definition incorporate elliptic anisotropy, see Appendix of
\cite{Liao:2008vj} for an example of such curved coordinates
which may be used to develop a similar embedding-reduction
procedure describing transverse elliptic flow. Another possibility
will be combining certain conformal transformations with
hydrodynamics equations to degrade the elliptic geometry back to
a spherical one.

{\em 2D Hubble embedding:} In all the previously discussed, we
have chosen to embed (1+1)D Hubble flow for the $(t,z)\to
(\tau,\eta)$ part, due to an emphasis on RHIC evolution.
Theoretically, one can also embed a (1+2)D Hubble flow for the
$(t,x,y)\to (t,\rho,\phi) \to (\tau_{\rho},\eta_{\rho},\phi)$
part, and can eventually reduce the equations to a velocity
equation with two variables $(\tau_\rho,z)$ in exactly the same
manner as before.

%\vspace{0.1in}

\begin{acknowledgements}
%\begin{ack}
The work is supported by the Director, Office of Energy Research,
Office of High Energy and Nuclear Physics, Divisions of Nuclear
Physics, of the U.S. Department of Energy under Contract No.
DE-AC02-05CH11231. JL is grateful to Edward Shuryak for very
helpful discussions.
%\end{ack}
\end{acknowledgements}

\section*{Appendix A}
\renewcommand{\theequation}{A\arabic{equation}}
\setcounter{equation}0

In this Appendix we give the detailed derivations leading to the
solutions in Section IV with the transverse velocity ansatz
(\ref{eqn_app_v}).

We first evaluate the derivatives $\frac{\partial {\mathcal
F}}{\partial \rho}$ and $\frac{\partial {\mathcal G}}{\partial
\tau}$ with ${\mathcal F},{\mathcal G}$ given in
(\ref{eqn_3d_f})(\ref{eqn_3d_g}). Again we will make use of $\bar
v_\rho\, _{,\, \rho}=\bar v_\rho \cdot C / \rho$ and $\bar
v_\rho\,  _{,\, \tau}= \bar v_\rho \cdot B / \tau$ for the
velocity ansatz (\ref{eqn_app_v}).

The result for $\frac{\partial {\mathcal F}}{\partial \rho}$ is
\begin{eqnarray}\label{eqn_3d_f_rho}
\frac{\partial {\mathcal F}}{\partial \rho} =&& \frac{\bar
v_\rho}{\nu (1-\nu)\rho^2 \tau^2 (1-\bar v_\rho^2)^3} \times
{\bigg \{ }  f_1 \rho\tau \bar v_\rho (1-\bar v_\rho^2) \nonumber \\
&& \, +f_2 \tau^2 \left[(1+C)\bar v_\rho^2+(C-1)\right](1-\bar
v_\rho^2)
\nonumber \\
&& \, +f_3\tau \rho \bar v_\rho + f_4 \tau^2 \left[(3C+1)\bar
v_\rho^2+(C-1)\right]
 {\bigg \}} \nonumber \\
 =&& \frac{\bar
v_\rho}{\nu (1-\nu)\rho^2 \tau^2 (1-\bar v_\rho^2)^3} \times
{\bigg \{ } \left[ -f_2 (C+1)\right]\tau^2 \bar v_\rho^4 \nonumber
\\
&& \, +\left[ -f_1 \right]\rho \tau \bar v_\rho^3 +
\left[2f_2+f_4(3C+1) \right]\tau^2 \bar v_\rho^2 \nonumber \\
&& \, +\left[ f_1+f_3 \right] \rho\tau \bar v_\rho + \left[
(f_2+f_4)(C-1)\right]\tau^2 {\bigg \}}
\end{eqnarray}
with coefficients $f_{1,2,3,4}$ given by
\begin{eqnarray}
&& f_1=-[B+(1-4B)\nu+2B\nu^2]\times (2C) \nonumber \\
&& f_2=-[C+(1-C)\nu] \nonumber \\
&& f_3=B\times (4C) \times (1-2\nu) \nonumber \\
&& f_4=C\times (1-2\nu)
\end{eqnarray}

The result for $\frac{\partial {\mathcal G}}{\partial \tau}$ is
\begin{eqnarray}\label{eqn_3d_g_tau}
\frac{\partial {\mathcal G}}{\partial \tau} =&& \frac{\bar
v_\rho}{\nu (1-\nu)\rho^2 \tau^2 (1-\bar v_\rho^2)^3} \times
{\bigg \{ }  g_1 \rho\tau \bar v_\rho (1-\bar v_\rho^2) \nonumber \\
&& \, +g_2 \rho^2 \left[(1+B)\bar v_\rho^2+(B-1)\right](1-\bar
v_\rho^2)
\nonumber \\
&& \, + g_3\tau \rho \bar v_\rho + g_4 \rho^2 \left[(3B+1)\bar
v_\rho^2+(B-1)\right]
 {\bigg \}} \nonumber \\
 =&& \frac{\bar
v_\rho}{\nu (1-\nu)\rho^2 \tau^2 (1-\bar v_\rho^2)^3} \times
{\bigg \{ } \left[ -g_2 (B+1)\right]\rho^2 \bar v_\rho^4 \nonumber
\\
&& \, +\left[ -g_1 \right]\rho \tau \bar v_\rho^3 +
\left[2g_2+g_4(3B+1) \right]\rho^2 \bar v_\rho^2 \nonumber \\
&& \, +\left[ g_1+g_3 \right] \rho\tau \bar v_\rho + \left[
(g_2+g_4)(B-1)\right]\rho^2 {\bigg \}}
\end{eqnarray}
with coefficients $g_{1,2,3,4}$ given by
\begin{eqnarray}
&& g_1=-[-C-\nu+2C\nu^2]\times (2B) \nonumber \\
&& g_2=-[(B-1)\nu] \nonumber \\
&& g_3= (4B) \times C \times (2\nu-1) \nonumber \\
&& g_4=B\times (2\nu-1)
\end{eqnarray}

Now combining the results into (\ref{eqn_v_hydro}) we obtain the
following (with $\bar v_\rho=A \tau^B \rho^C$ substituted in)
\begin{eqnarray}\label{eqn_3d_power_law}
\frac{\partial {\mathcal F}}{\partial \rho}-\frac{\partial
{\mathcal G}}{\partial \tau} = && \frac{\bar v_\rho}{\nu
(1-\nu)\rho^2 \tau^2 (1-\bar v_\rho^2)^3} \times {\mathcal I}
[\tau,\rho]  \mathbf{=0}
\nonumber \\
\mathbf{0=}\,{\mathcal I}[\tau,\rho] = && \,\, [-f_2(C+1)A^4]\tau^{4B+2}\rho^{4C} \nonumber \\
&& + [g_2(B+1)A^4]\tau^{4B}\rho^{4C+2} \nonumber \\
&& + [(g_1-f_1)A^3]\tau^{3B+1}\rho^{3C+1} \nonumber \\
&& + [(2f_2+f_4(3C+1))A^2]\tau^{2B+2}\rho^{2C} \nonumber \\
&& + [(-2g_2-g_4(3B+1))A^2]\tau^{2B}\rho^{2C+2} \nonumber \\
&& + [(f_1+f_3-g_1-g_3)A]\tau^{B+1}\rho^{C+1}   \nonumber \\
&& + [(f_2+f_4)(C-1)]\tau^2 \nonumber \\
&& + [-(g_2+g_4)(B-1)]\rho^2
\end{eqnarray}
Again one can test the correctness of the above equation by using
the Hwa-Bjorken ($A=0$) and the 3D Hubble ($A=1,B=-1,C=1$)
solutions.

In  equation (\ref{eqn_3d_power_law}), terms with
various powers of $\tau,\rho$ (and only power terms) appear in
${\mathcal I}[\tau,\rho]$: to make all of them, either mutually
cancel (among terms with exactly the same $\tau,\rho$ powers) or
vanish by respective coefficients, to eventually zero is quite
nontrivial. A thorough sorting of the sequences of $\tau,\rho$
powers can exhaust all possibilities to satisfy the algebraic
equation (\ref{eqn_3d_power_law}).

To see how this actually works, we give one concrete example.
Let's consider the case when $B>0$ and $C\ne 0$: this implies that
for the exponents of $\tau$ we have $4B+2>4B>2B$, $4B+2>3B+1>B+1$,
$4B+2>2B+2>2B$, $2B+2>2$ and $2B+2>B+1$. So the term
$[-f_2(C+1)A^4]\tau^{4B+2}\rho^{4C}$ in
Eq.(\ref{eqn_3d_power_law}) can NOT be cancelled by any other one
and has to vanish by itself: this leads to
\begin{equation}
f_2(C+1)A^4=0
\end{equation}
which in turn gives three possibilities $f_2=0$ or $C=-1$ or
$A=0$. In this example we follow $C=-1$ (other choices lead to
other solutions). With $C=-1$ we notice again that the term
$[-(g_2+g_4)(B-1)]\rho^2$ can NOT by cancelled by any other
remaining terms and thus shall vanish by itself: this leads to
\begin{equation}
-(g_2+g_4)(B-1)=0
\end{equation}
which again has two possibilities $B=1$ or $g_2+g_4=0$. Now we
choose to follow $B=1$: with this choice the remaining terms are
significantly simplified and finally lead to two equations about
the coefficients:
\begin{eqnarray}
&& 2 g_2 A^4 + (g_1-f_1) A^3 + 2(f_2-f_4)A^2=0 \nonumber \\
&& -2(g_2+2g_4)A^2 + (f_1+f_3-g_1-g_3) A -2 (f_2+f_4)=0 \nonumber
\\
\end{eqnarray}
It can then be verified that the only solution is $A=1$ for
arbitrary $\nu$.

Of course there are many but finite number of combinations that
one can follow to check one by one. Note not all possibilities
appearing initially can finally lead to a solution: there are only
four variables $A,B,C,\nu$ and in most cases it turns out
contradiction occurs at the end which means no solution. After a
tedious examination we have found all possible solutions as listed
in Section IV, and there is no more solution of the power law
ansatz type as in (\ref{eqn_app_v}).

\section*{Appendix B}
\renewcommand{\theequation}{B\arabic{equation}}
\setcounter{equation}0

In this Appendix we use (1+1)D ideal relativistic hydrodynamics to
demonstrate the reduction method in a more explicit manner. The
hydrodynamics equations are (in (t,z) coordinates)
\begin{eqnarray}
\frac{[\partial_t+v\partial_z]\epsilon}{\epsilon+p} &=&
-\partial_z v - \gamma_v^2
[\partial_t+v\partial_z] (\frac{v^2}{2}) \\
\gamma_v^2 [\partial_t+v\partial_z] v &=& -\frac{\partial_z
p}{\epsilon+p}-\frac{v\partial_t p}{\epsilon+p}
\end{eqnarray}
In the above $v$ is the spatial velocity $dz/dt$ and
$\gamma_v\equiv 1/\sqrt{1-v^2}$. The energy density $\epsilon$ and
pressure $p$ shall be related by the E.o.S, which we use in a
slightly different way. We introduce the enthalpy density
$w=\epsilon+p$ and the speed of sound $c_s\equiv \sqrt{\partial p
/ \partial \epsilon}$ (which can be deduced from E.o.S), and use
the following relations
\begin{equation}
d\epsilon=\frac{1}{1+c_s^2}dw \,\, , \,\,
dp=\frac{c_s^2}{1+c_s^2}dw
\end{equation}
to re-write the hydrodynamics equations into:
\begin{eqnarray}
\partial_t [\mathbf{ln}(w)] + v\, \partial_z [\mathbf{ln}(w)] &=&
-\frac{\gamma_v^2}{1-\xi} [\partial_x v + v\, \partial_t v] \quad
\\
v\, \partial_t [\mathbf{ln}(w)] + \partial_z [\mathbf{ln}(w)] &=&
-\frac{\gamma_v^2}{\xi} [v\, \partial_x v + \partial_t v]
\end{eqnarray}
with $\xi \equiv c_s^2/(1+c_s^2)$. From these two equations we can
obtain $\partial_t [\mathbf{ln}(w)]$ and $\partial_z
[\mathbf{ln}(w)]$:
\begin{eqnarray}
&& \partial_t [\mathbf{ln}(w)] = {\mathcal X}[v,\partial_t v,\partial_z v]={\mathcal X}[t,z]\nonumber \\
&&\,\, = \frac{-\gamma_v^4}{\xi (1-\xi)} {\bigg \{ } {\big [}
\xi-(1-\xi)v^2 {\big ]}\partial_z v + {\big [ } 2\xi-1 {\big ]}
v\,
\partial_t v {\bigg \}
} \nonumber \\
\\
&& \partial_z [\mathbf{ln}(w)] = {\mathcal Y}[v,\partial_t v,\partial_z v]={\mathcal Y}[t,z]\nonumber \\
&&\,\, = \frac{-\gamma_v^4}{\xi (1-\xi)} {\bigg \{ } {\big [}
(1-\xi)-\xi v^2 {\big ]} \partial_t v + {\big [} 1-2\xi {\big ]}
v\,
\partial_z v {\bigg \}
} \nonumber \\
\end{eqnarray}
The necessary and sufficient conditions for the above set of
equations to be soluble is the following:
\begin{eqnarray} \label{eqn_1d_reduced}
\partial_z {\mathcal X}[t,z] - \partial_t {\mathcal Y}[t,z]=0
\end{eqnarray}
Thus we have reduced the original hydrodynamics equations into a
single but second order differential equation for the velocity
field $v(t,z)$ only. With the above satisfied, the matter field is
given by
\begin{equation} \label{eqn_1d_matter}
w(t,z)=w_0 \cdot e^{\int_{t_0}^{t} dt' {\mathcal
X}[t',z]+\int_{z_0}^{z} dz' {\mathcal Y}[t,z']}
\end{equation}
with $w_0$ its value at arbitrary reference point $(t_0,z_0)$.

In the case of a linear E.o.S as the one in (\ref{eqn_eos}), the
speed of sound $c_s$ and thus $\xi$ are constants independent of
$\epsilon$ or $p$, and we can further simplify the reduced
equation (\ref{eqn_1d_reduced}) for velocity field into the
following:
\begin{eqnarray} \label{eqn_1d_simple}
&& [(1-\xi)-\xi v^2](1-v^2)(\partial_t \partial_t v) \nonumber \\
&&\, +[(2-3\xi)-\xi
v^2] (2v) (\partial_t v)^2 \nonumber \\
&& \, - [\xi-(1-\xi)v^2](1-v^2)(\partial_z \partial_z v)
\nonumber \\
&& \, -[(3\xi-1)+(\xi-1)v^2](2v) (\partial_z v)^2 \, \nonumber \\
&& \, +2(1-2\xi)(1-v^2)v (\partial_t \partial_z v)  \nonumber \\
&& \, +2(1-2\xi)(1+3v^2) (\partial_t v)(
\partial_z v) = \, 0
\end{eqnarray}
A similar scheme can be carried out for curved coordinates like
$(\tau,\eta)$ in a straightforward way. We notice a similar
implementation using light-cone variables $z_{\pm}=t\pm z$ in
\cite{Bialas:2007iu}.

We emphasize that while the above procedure seems somewhat trivial
in (1+1)D, its realization is much more nontrivial and involved in
(1+3)D. We also point out that the reduced equation
(\ref{eqn_1d_reduced}) for the velocity field (or the simplified
one in case of linear E.o.S) has to be satisfied by {\em all}
solutions to the (1+1)D hydrodynamics equations. In Appendix C we
give a nontrivial and involved example from the recently found
Nagy-Csorgo-Csanad (NCC) solutions\cite{Nagy:2007xn} (which also
include (1+1)D Hwa-Bjroken as a special case) to show the
correctness and usefulness of the derived velocity equation.

\section*{Appendix C}
\renewcommand{\theequation}{C\arabic{equation}}
\setcounter{equation}0

In this Appendix we show that the NCC family of analytic solutions
in \cite{Nagy:2007xn} for 1-D ideal hydrodynamics equations with a
linear E.o.S can also be deduced by subjecting their velocity
field ansatz to the reduced equations (\ref{eqn_1d_simple}) we
derived. With the resulting velocity field we also show the matter
field of NCC solutions is indeed given by (\ref{eqn_1d_matter}).

The velocity field ansatz of NCC solutions is the following (for
inside-light-cone region, i.e. $|z|<|t|$):
\begin{equation}
v=\tanh[\lambda
\eta]=\frac{(t+z)^\lambda-(t-z)^\lambda}{(t+z)^\lambda+(t-z)^\lambda}
\,\, , \,\, \eta=\frac{1}{2}\mathbf{ln}[\frac{t+z}{t-z}]
\end{equation}
with $\lambda$ some constant. With the above, we obtain the
following relations for the derivatives:
\begin{eqnarray}
&& \partial_t v=\lambda (1-v^2) \partial_t \eta \,\,\, , \,\,\,
\partial_z v=\lambda (1-v^2) \partial_z \eta \nonumber \\
&& \partial_t\partial_t v =\lambda (1-v^2)[(\partial_t \partial_t
\eta) - 2\lambda v (\partial_t \eta)^2] \nonumber \\
&& \partial_z\partial_z v =\lambda (1-v^2)[(\partial_z \partial_z
\eta) - 2\lambda v (\partial_z \eta)^2] \nonumber \\
&& \partial_t\partial_z v =\lambda (1-v^2)[(\partial_t \partial_z
\eta) - 2\lambda v (\partial_t \eta)(\partial_z \eta)]
\end{eqnarray}
Substituting the above derivatives into our velocity equation
(\ref{eqn_1d_simple}), we obtain the following:
\begin{eqnarray}
&& 0 = \lambda (1-v^2)^2 \times \nonumber \\
&&\,\,  {\bigg \{ } [(1-\xi)-\xi v^2](\partial_t\partial_t \eta) +
[-\xi+(1-\xi)v^2](\partial_z\partial_z \eta)  \nonumber \\
&&\,\,  +2(1-2\xi)v(\partial_t\partial_z \eta) + 2(1-2\xi)\lambda
(1+v^2)(\partial_t \eta)( \partial_z \eta) \nonumber \\
&&\,\,  + 2(1-2\xi)\lambda v[(\partial_t \eta)^2+(\partial_z
\eta)^2]
  {\bigg \} }
\end{eqnarray}
After evaluating the derivatives of $\eta$ in the above, we
obtain:
\begin{equation}
\frac{2\lambda (1-v^2)^2}{(t^2-z^2)^2} \cdot (1-2\xi) \cdot
(1-\lambda)\cdot [(tz)v^2-(t^2+z^2)v+(tz)]=0
\end{equation}
We find three classes of solutions:
\begin{itemize}
\item $\xi=1/2$ with arbitrary $\lambda$; \\
\item $\lambda=1$ with arbitrary $\xi$ (which is nothing but the Hwa-Bjorken solution); \\
\item $[(tz)v^2-(t^2+z^2)v+(tz)]=0$ which yields only one
meaningful solution with $v=z/t$, but this is just the $\lambda=1$
solution.
\end{itemize}
These cover all (1+1)D NCC solutions found in \cite{Nagy:2007xn}.
Note their parameter $\kappa$ from E.o.S $\epsilon=\kappa p$ is
related to our E.o.S parameter $\xi \equiv c_s^2/(1+c_s^2)$ by
$\kappa=(1-\xi)/\xi$.

Next we examine the matter field corresponding to the velocity
field solutions:
\begin{itemize}
\item for $\xi=1/2$ case: we have ${\mathcal
X}=(-\lambda)[2t/(t^2-z^2)]$ and ${\mathcal
Y}=(-\lambda)[-2z/(t^2-z^2)]$, which via our equation
(\ref{eqn_1d_matter}) gives
$w=w_0 {\big[}\frac{t_0^2-z_0^2}{t^2-z^2}{\big ]}^{\lambda} $; \\
\item for $\lambda=1$ case: we have $v=z/t$ and thus ${\mathcal
X}=\frac{-1}{2(\xi-1)}[2t/(t^2-z^2)]$ and ${\mathcal
Y}=\frac{-1}{2(\xi-1)}[-2z/(t^2-z^2)]$, which via our equation
(\ref{eqn_1d_matter}) gives $w=w_0
{\big[}\frac{t_0^2-z_0^2}{t^2-z^2}{\big ]}^{1/(2-2\xi)} $
\end{itemize}
The two cases can be combined into a single form:
\begin{equation}
w=w_0 \times {\big[}\frac{t_0^2-z_0^2}{t^2-z^2}{\big
]}^{\frac{\lambda}{2(1-\xi)}}
\end{equation}
which is the same as obtained in \cite{Nagy:2007xn}.

\end{document}